# Influence of Slope Angle on Deposit Morphology and Propagation of Laboratory Landslides

Yan-Bin Wu[1] Zhao Duan[2*, 3] Jian-Bing Peng[4,5] Qing Zhang[2, 3] Thomas Pähtz[1*,6]

[1] Ocean College, Zhejiang University, Zhoushan 316021, China
[2] College of Geology and Environment, Xi'an University of Science and Technology, Xi'an 710054, China
[3] Shaanxi Provincial Key Laboratory of Geological Support for Coal Green Exploitation, Xi'an 710055, China
*Corresponding author: +86 15991609977; email: duanzhao@xust.edu.cn; address: 58 Yanta Middle Road, Xi'an 710054, China
[4] School of geology engineering and Geomatics, Chang'an University, Xi'an 710054, China
[5] Key Laboratory of western Mineral Resources and Geological Engineering of Ministry of Education, Chang'an University, Xi'an 710054, China
[6] Donghai Lab, Zhoushan 316021. China
*Corresponding author: +86 13148327827; email: tpaehtz@gmail.com; address: Zhoushan 316021, China

**Abstract**

Landslide deposits often exhibit surface features, such as transverse ridges and X-shaped conjugate troughs, whose physical formation origins are not well understood. To study the deposit morphology, laboratory studies typically focus on the simplest landslide geometry: an inclined plane accelerating the sliding mass immediately followed by its deceleration on a horizontal plane. However, existing experiments have been conducted only for a limited range of the slope angle $\theta$. Here, we study the effect of $\theta$ on the kinematics and deposit morphology of laboratory landslides along a low-friction base, measured using an advanced 3D scanner. At low $\theta$ (30°-35°), we find transverse ridges formed by overthrusting on the landslide deposits. At moderate $\theta$ (40°-55°), conjugate troughs form. A Mohr-Coulomb failure model predicts the angle enclosed by the X-shaped troughs as 90°-$\varphi$, with $\varphi$ the internal friction angle, in agreement with our experiments and a natural landslide. This supports the speculation that conjugate troughs form due to failure associated with a triaxial shear stress. At high $\theta$ (60°-85°), a double-upheaval morphology forms because the rear of the sliding mass impacts the front during the transition from the slope to the horizontal plane. The overall surface area of the landslides increases during their downslope motion and then decreases during their runout.

## 1 Introduction

Landslides can be very destructive, especially when they runout over large distances due to high mobility [1-7]. Apart from field investigations, one can study their flowing behaviour via building physical models of simplified landslide geometries and carrying out laboratory experiments on them [8-12]. Of particular interest is a landslide's deposit morphology, since it conveys information about the granular processes that have been at work during its slide.

Previous physical-model experiments [13-20] and field investigations [16,17,21-23] have revealed different deposit morphologies of landslides and their physical origin. For example, levee formation has been linked to static zones near the lateral boundaries of unconfined dry granular flows [24]. There is also widespread agreement that commonly occurring transverse ridges, forming perpendicular to the flow direction, are compression-related surface features [17,18,21]. However, the physical origin of conjugate troughs (i.e., surface structures with a characteristic X-shape), observed on the surface of some large-landslide deposits, is less clear. Based on field investigations, Wang, et al. [21] and Zhao, et al. [25] speculated that they form by the interplay between transport-parallel compression and radial or lateral spreading during a landslide's runout, the latter giving rise to a triaxial shear stress. If this speculation was true, it would imply that the degree of the initial landslide acceleration plays a crucial role in the formation process, since compression during a landslide's runout is the result of its sudden deceleration during its transition from the initial slope to the much flatter runout terrain. This in turn suggests that the initial-slope angle is a key parameter controlling the occurrence of conjugate troughs. It is one of the objectives of this paper to test this hypothesis by means of physical-model experiments.

While numerous previous laboratory studies investigated inclined-plane geometries [10,26-33], only a few investigated landslide geometries, that is, a sudden [17,18,34] or smooth [12,16] transition from an inclined plane to much flatter runout terrain. However, most of the latter studies focused on the landslide dynamics rather than the deposit morphology. The only exception is Shea and van Wyk de Vries [16], who studied only the deposit morphology, though they did not identify conjugate troughs. Furthermore, all previous laboratory studies based on a landslide geometry did not consider a large range of slope angles.

Here, we conduct laboratory experiments based on a landslide geometry, with slope angles varying between 30° and 85°. Note that granular flows with slope angles between 70° and 90° do in fact occur in nature, such as in cliff avalanche events and chalk flows in coastal areas [19,35]. To record the landslide evolution, we use two high-speed cameras and an advanced 3D scanner.

The focus of our experiments lies on both the landslide dynamics and deposit morphology, especially conjugate troughs. Our objectives are as follows: 1) explore variations in landslide motion parameters and states with slope angle; 2) determine the influence of slope angle on deposit morphology and identify the physical mechanisms behind the formation of surface features, especially conjugate troughs; 3) explore the temporal evolution of the sliding masses' length, width, and area during their entire motion.

## 2 Methods

### 2.1 Experimental setup

A sandbox experiment is performed to study the motion process and deposit morphology of laboratory landslides (Figure 1). Plexiglass is used to construct the experimental devices, which were composed of five parts: an inclined plate, a horizontal plate, a sand container, a 3D scanner, and two high-speed cameras. A pair of sandbox tracks is put on the inclined plate to adjust the height of the sandbox. The lengths of the inclined plate and horizontal plate are 1.5 m and their widths are 1.2 m. The slope angle can be varied between 30° and 90°. A coordinate system is defined in Figure 1, in which $x$ denotes the direction of a landslide's mean motion and $z$ the vertical direction oriented upwards. The fixed volume of the sandbox is $3.6 \times 10^{-3} \mathrm{m}^3$. It consists of a gate that can be rapidly opened to release the sand. A three-dimensional (3D) scanner (F6 Smart, MANTIS VISIONS) operates at 8 frames/s and 1.3-megapixel resolution. It obtains 3D coordinate data of the upper surface with an accuracy of 0.1 mm during the whole landslide motion process. It consists of three lenses: one at the bottom that emits near-infrared (NIR) light towards the sliding mass and two lenses at the top, one receiving the back-reflected NIR light and one that can produce coloured images. The received NIR data are transformed into 3D cloud data of the surface morphology. The 3D data are produced according to the principles of stereoscopic parallax and active triangular ranging [13,14,36]. Two high-speed cameras (60 frames/s, 0.4-megapixel resolution) are used to collect images at the end of each experiment. One is placed on a moveable camera shelf, which allows taking deposit photos from a bird view. The other one is fixed at the front of the horizontal plate, with a horizontal view.

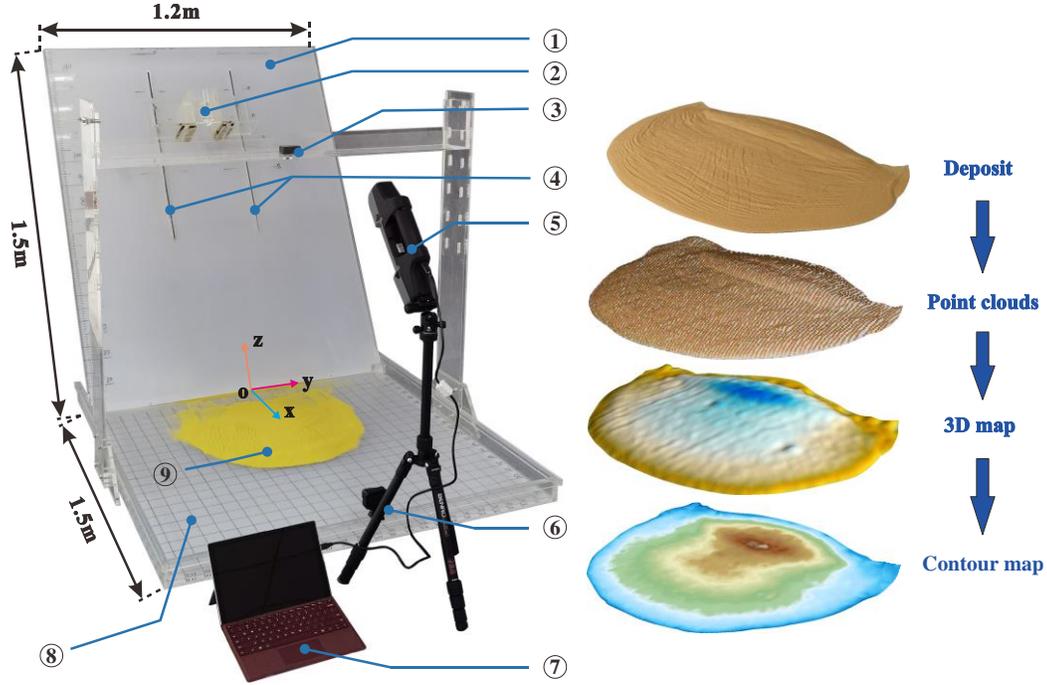

① Inclined plate   ② Sandbox   ③ High-speed camera 1   ④ Track of sandbox   ⑤ 3D scanner
⑥ High-speed camera 2   ⑦ Data acquisition   ⑧ Horizontal plate   ⑨ Deposits

Figure 1 Apparatus used for the physical-model experiments.

## 2.2 Material

Medium-fine quartz sand (inset of Figure 2) is used as the landslide material. Its particle size distribution (Figure 2) exhibits a mean of 0.18 mm, an uneven coefficient of $C_u = D_{60} / D_{10} = 2.39$, and a curvature coefficient of $C_c = D_{50}^2 / (D_{60} D_{10}) = 1.19$, where $D_n$ denotes the size that $n$% of particles do not exceed. The particles' surface area per unit mass is 0.02 m$^2$·kg$^{-1}$, and the sand's internal friction angle $\varphi$ is 36°, measured by direct shear tests [37,38]. The friction coefficient of the interface between the plate and the sand is 0.42.

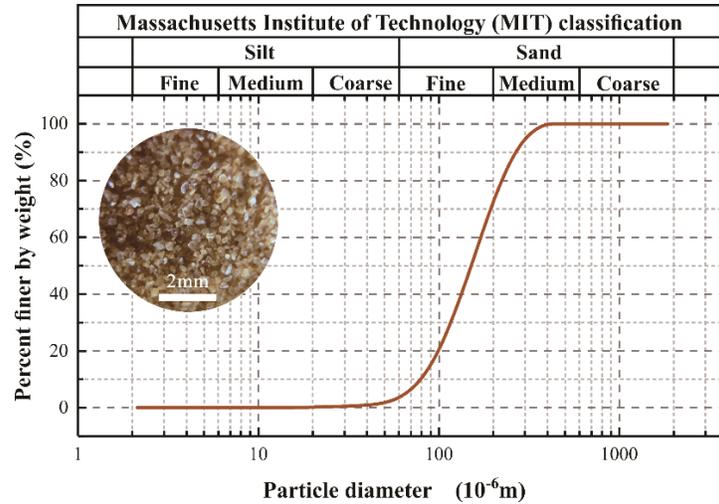

Figure 2 Particle size distribution of the experimental material (inset: image of the medium-fine quartz sand).

## 2.3 Experimental design

Before each experiment, the inclined plate, horizontal plate, and interior of the sandbox are wiped with an electrostatic-proof liquid. After the liquid dried, the sand is filled in the sandbox in three separate steps, interrupted by compaction to ensure uniformity. The sandbox is then closed by a gate that can be rapidly opened at the start of each experiment (Figure 1). The completely filled sandbox, containing $3.6 \times 10^{-3}$ m$^3$ of sand volume with a mass of 5.4 kg, is placed such that its centre-of-mass initial height is at 0.7 m in each experiment, by adjusting a pair of tracks (Figure 1).

Seven calibration tests on a slope of 50° are performed to quantify the random errors in the experiments. Then, the actual experiments are conducted for slope angles varying from 30° to 85° at intervals of 5°. All experiments are run at least twice and the morphology parameters, i.e., maximum deposit length, width, and depth, deposit area, length-width ratio, and circumference-area ratio, noted (Figure 3). If any of these parameters' difference between both runs relative to the mean of both runs is larger than the respective value for the calibration experiments (Figure 4), a third experiment is performed. Then, those two of the three experiments are selected that exhibit the smallest difference between one another.

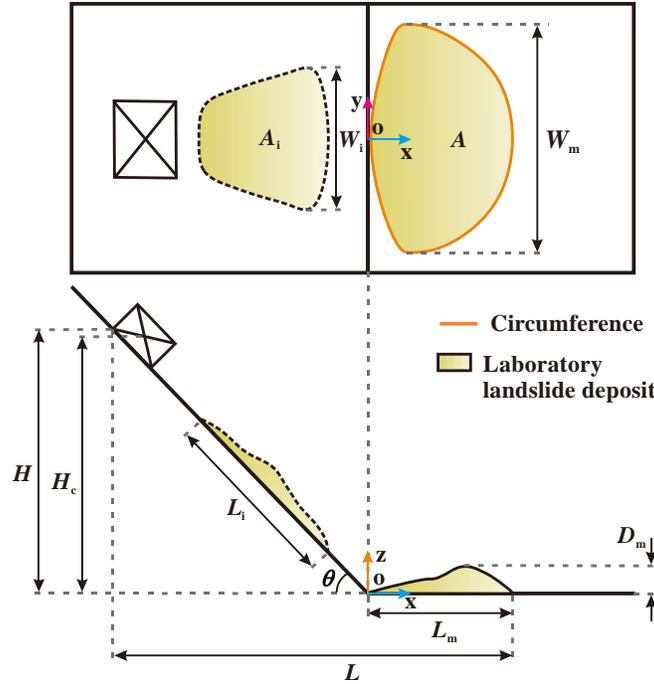

Figure 3 Diagram of the deposit morphology of our laboratory landslides: $L_m$ = maximum length (the projected length from the most rear part to the front part on the $x$-o-$y$ plane); $W_m$ = maximum width; $D_m$ = maximum depth; $A$ = area projected on horizontal plane; $C$ = circumference of deposit; $\theta$ = slope angle; $H$ = height of sandbox scarp; $H_c$ = height of sandbox centre of mass; $L$ = runout distance; $L_i$, $W_i$, $A_i$ = length, width, and area of sliding mass during its motion.

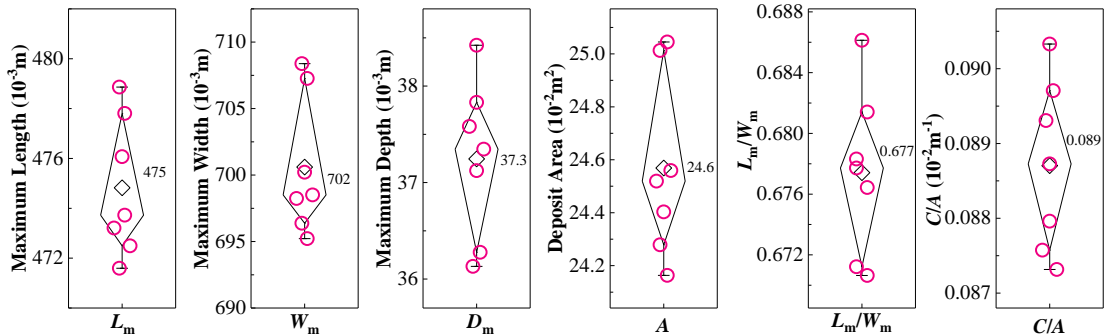

Figure 4 Calibration experiments. A circle corresponds to a measurement for a given run and the diamonds to the mean over all circles. The black lines roughly visualize the distribution.

## 3 Results

### 3.1 Motion characteristics

#### 3.1.1 Qualitative morphology patterns during motion and runout

The morphology of the sliding masses varies with time and slope angle. However, the overall qualitative behaviour tends to be similar for each of the slope angle intervals 30°-35°, 40°-55°, and 60°-85°. Therefore, snapshots of landslides at 30°, 50°, and 80° are shown as representative cases (Figure 5). At 30°, the landslide propagates as a thin and relatively uniform mass of nearly constant width and leaves a deposit on the inclined plate (Figure 5 (a)). At 50°, the sliding mass laterally spreads whilst propagating downslope, like a fan. Its thickness profile when propagating on the inclined plate is uneven, with clearly visible bumps around the centreline and less sand at the flanks. However, it leaves nearly no deposit on the inclined plate (Figure 5 (b)). At 80°, the fan-shaped expansion on the inclined plate lessens as the sliding mass falls almost freely. Its thickness profile on the inclined plate is very uneven (Figure 5 (c)). Moreover, a thinly spread layer forms in front of the main deposit (not accounted for when measuring runouts) due to a secondary impact of sliding mass from the rear [39,40].

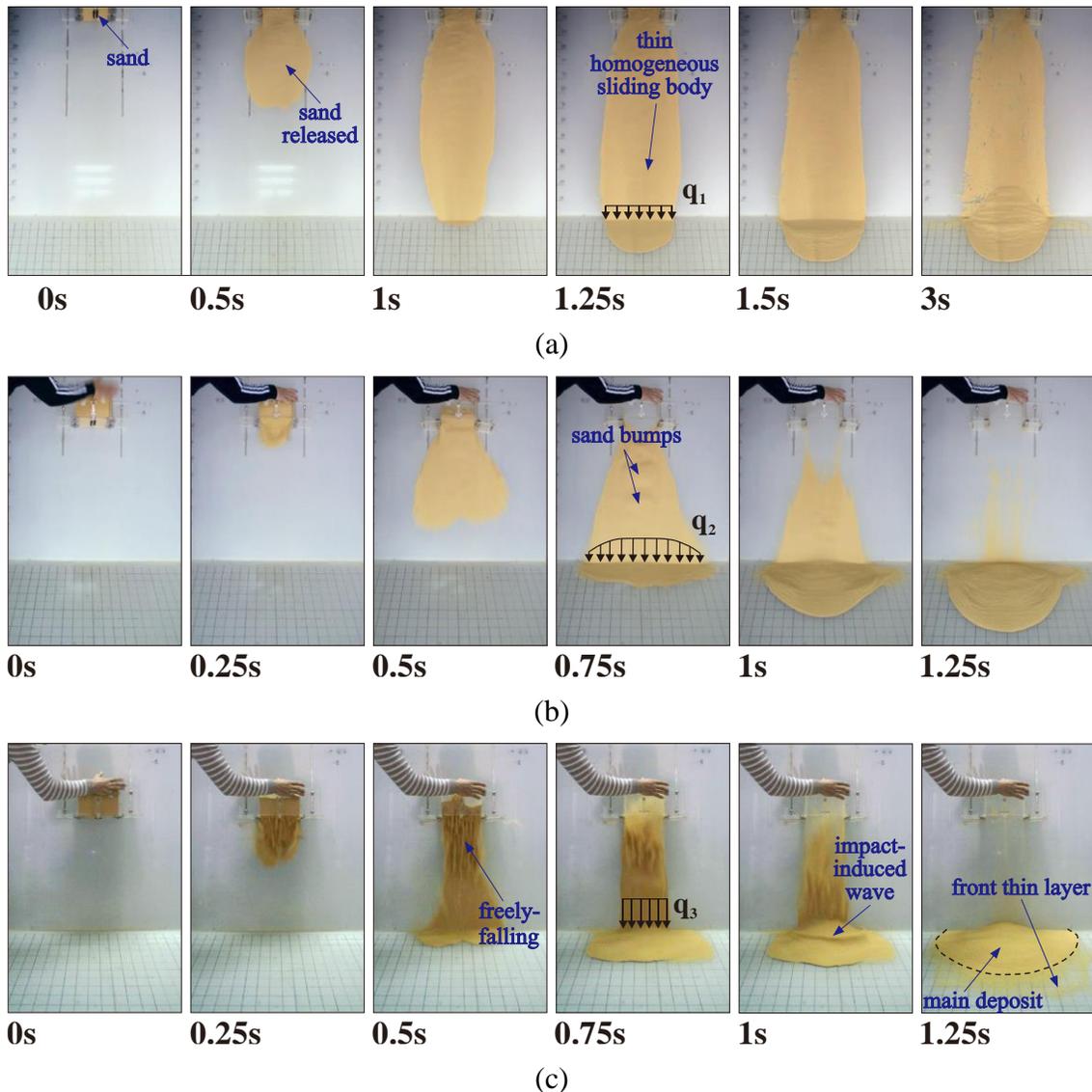

Figure 5 Motion processes of the laboratory landslides at slope angles of (a) 30°; (b) 50° and (c) 80°.

The runout ($L$, defined as in Figure 3) decreases linearly with slope angle $\theta$: $L = (-19.58\theta / ° + 2103.49)$ mm (Figure 6).

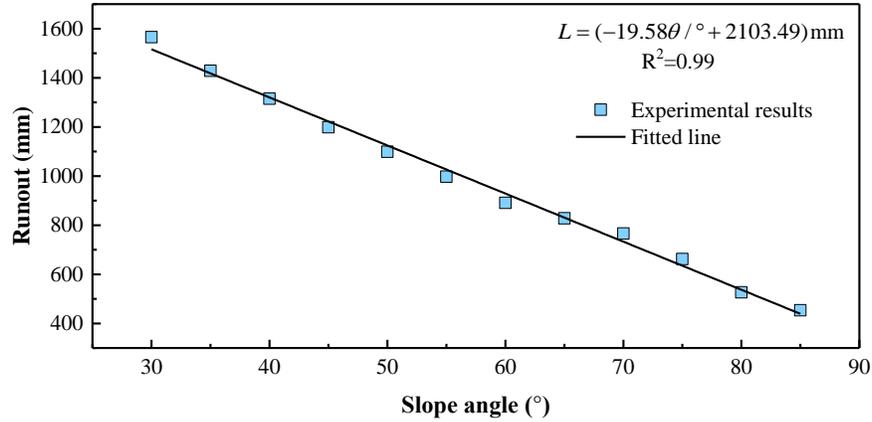

Figure 6 Runout of the laboratory landslides at different slope angles.

### 3.1.2 Dynamic parameters

The displacement of a sliding mass is defined as the difference between its front and starting position, which is the bottom of the sandbox. Its first and second derivative after time are its velocity and acceleration, respectively. Furthermore, the landslide front propagation duration is defined based on the instant the front of the sliding mass stops moving. All four dynamic parameters are shown in Figure 7. At 30°-35°, the landslide exhibits three stages: uniform acceleration, constant-velocity, and deceleration, consistent with previous laboratory studies at low slope angles [10,41,42]. The first two stages are before the sliding mass encounters the slope break (indicated by arrows and vertical dashed lines). At 40°-55°, the acceleration stage can be divided into a uniform acceleration stage and, less pronounced but still discernible, an acceleration stage at a decreasing rate. At 60°-85°, the landslides reach their peak velocity during the uniform acceleration stage and then almost immediately enter a deceleration stage when their fronts encounter the slope break. However, at 80°-85°, a brief secondary acceleration episode occurs during the deceleration stage. This phenomenon is closely related to the waves highlighted in Figure 5(c), which form due to the impact of the rear portion of the sliding mass on the deposit that has already accumulated on the horizontal plate. Then, this rear portion will leapfrog over the main deposit and form the thinly spread secondary frontal layer.

In general, alternating stages of acceleration and deceleration are signatures of stress fluctuations and were already previously observed by Roche, et al. [43] and Bachelet, et al. [44].

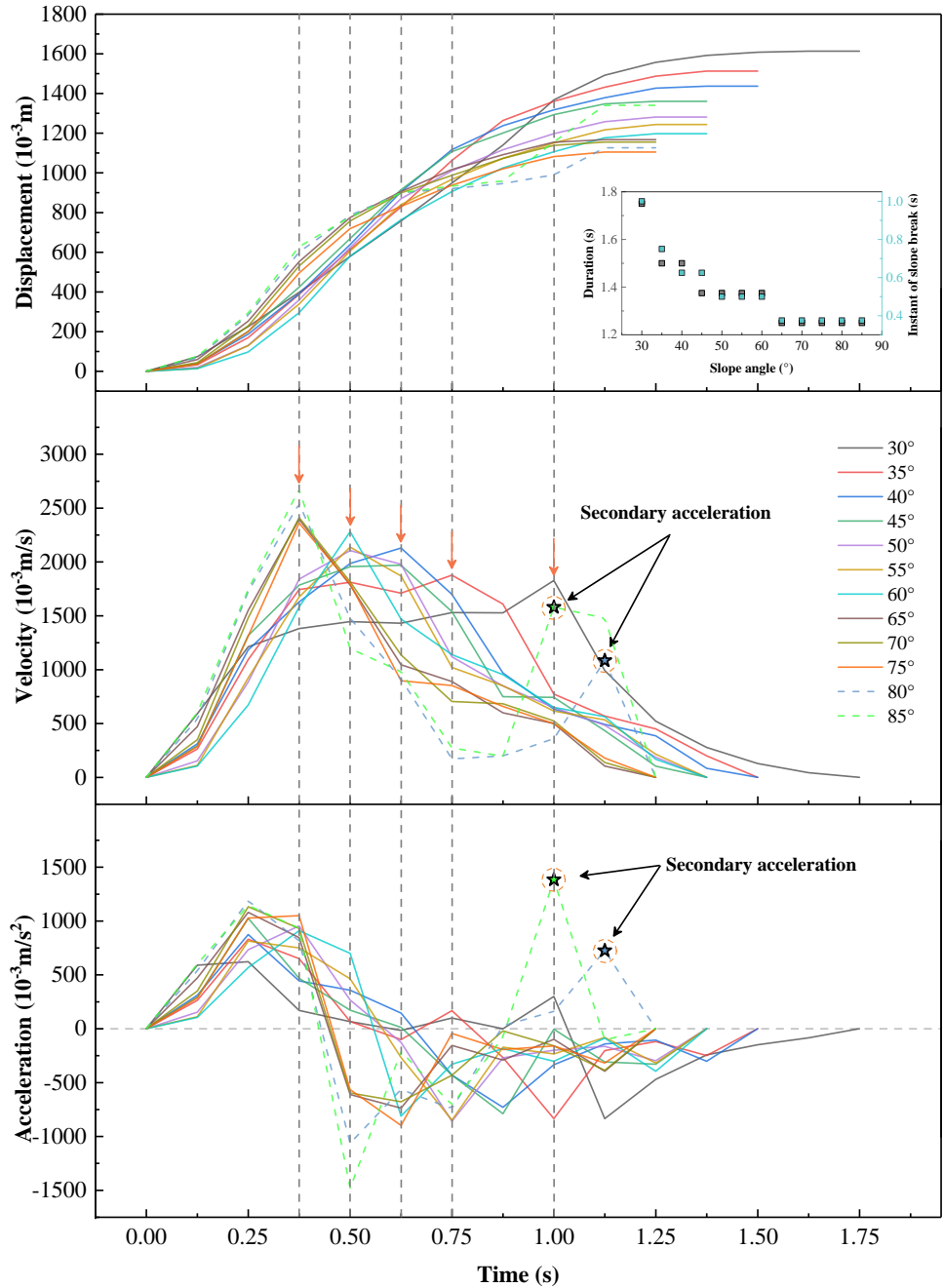

Figure 7 Time evolution of dynamic parameters for different slope angles. The orange arrows and vertical dashed lines indicate the instant a sliding mass arrives at the slope break (inset: the landslides' front duration and the moment they reach to the slope break).

### 3.1.3 Morphology parameters during the whole motion

The length, width, and area of a sliding mass are defined as shown in Figure 3. Both the length and area tend to increase less and less rapidly with time and eventually even decrease (Figure 8), though there is almost no such decrease in area for large slope angles $\theta > \approx 60°$. During the increase, the rate of change of length or area positively correlates with the slope angle, whereas the maximum length or area exhibits a negative correlation. Interestingly, the maximum length and area values for $\theta > \approx 60°$ have almost no dependency on the slope angle.

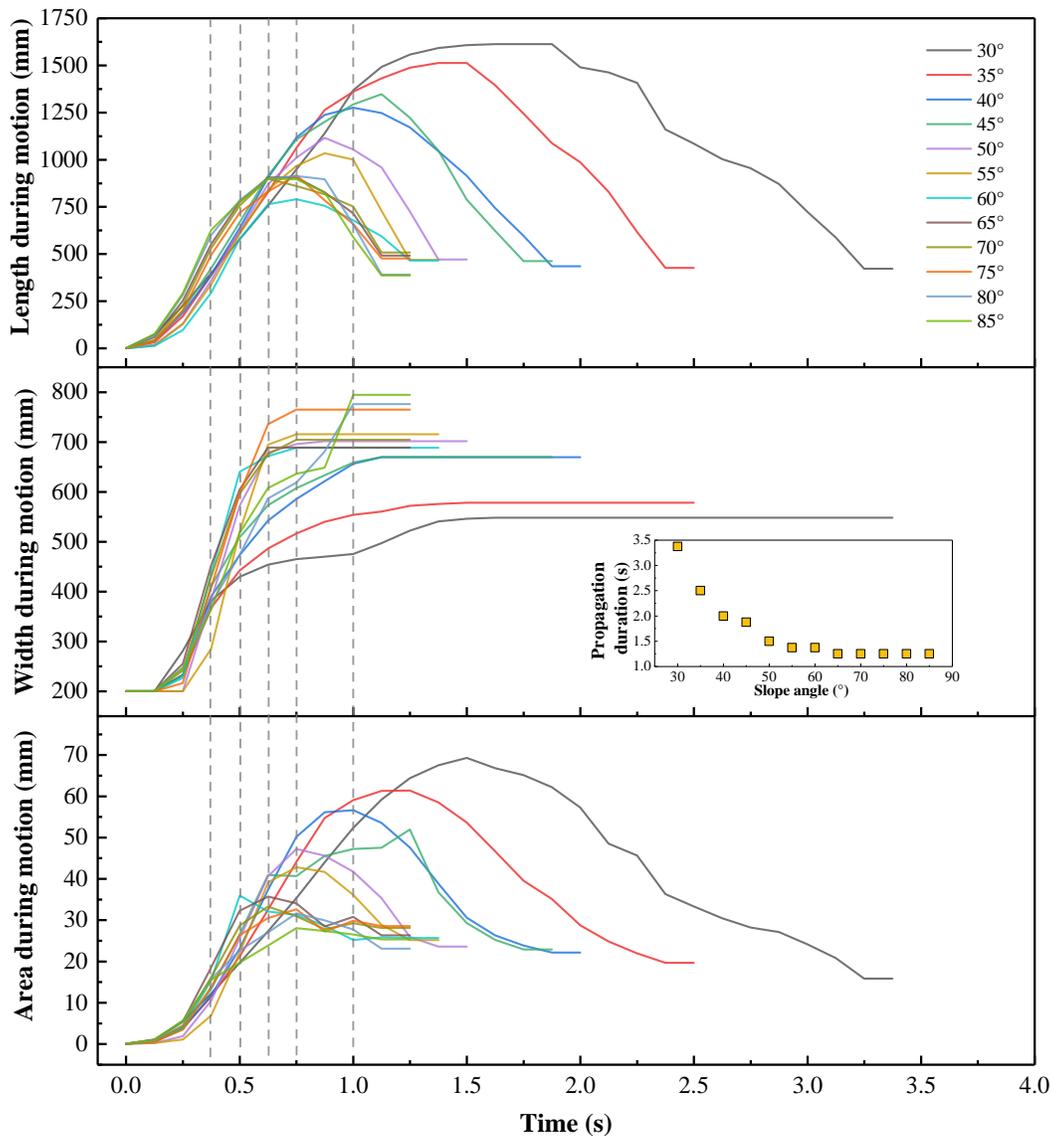

Figure 8 Time evolution of morphological parameters at different slope angles. Dashed lines indicate the landslide's front reaches to slope breaks (cf. Figure 7).

The time evolution of the width of the sliding mass undergoes three distinct stages. In the first stage, the width is about equal to the width of the sandbox. In the second stage, the width increases to its maximum (namely, the maximum deposit width), with a rapidly increasing rate at 50°-75°, but with a relatively low increasing at the other slope angles. In the third stage, the sliding mass moves forward without significantly changing its maximum width.

The inset of Figure 8 depicts the relation between the propagation duration of the landslides, defined based on the instant the whole motion stops everywhere (as opposed to only the front in the inset of Figure 7), and the slope angle.

The deposit parameters of the landslides as functions of the slope angle are shown in Figure 9.

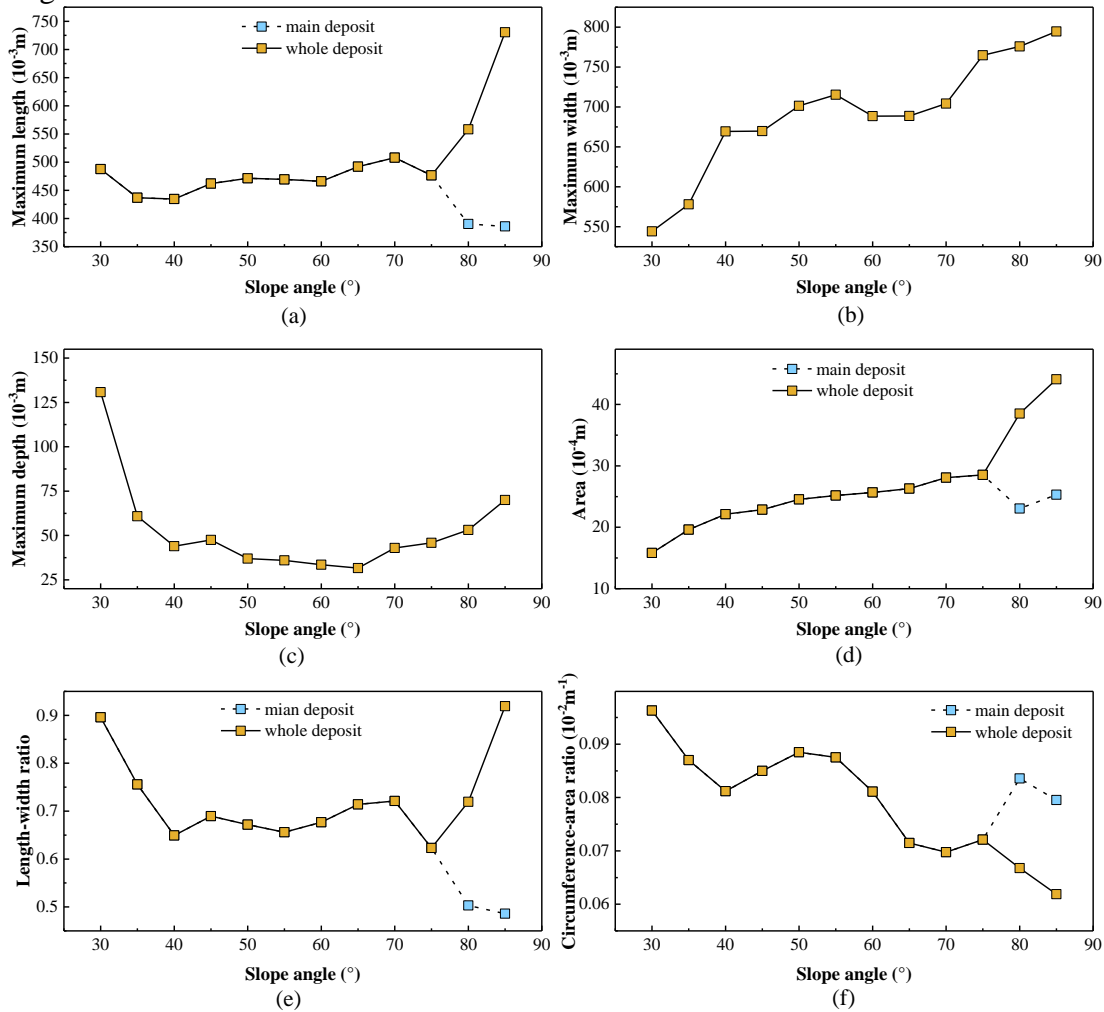

Figure 9 Deposit morphology parameters of laboratory landslides, as defined in Figure 3: (a) maximum length, (b) maximum width, (c) maximum depth, (d) area, (e) length-width ratio; and (f) circumference-area ratio. At 80°-85°, a secondary front consisting of thinly spread deposit exist in addition to the main deposit (cf. Figures 5(c), 10(c), and 10(d)).

## 3.2 Deposit morphology

### 3.2.1 Deposit surface features

Transverse ridges are widely developed on the surfaces of the deposits at 30°-35° (Figure 10(a)). Those on the deposit centres are oriented perpendicular to the landslides' mean motion direction ($x$ axis direction) and those on the deposit flanks exhibit an angle to the $x$-direction. The latter are more densely distributed than the former, not only at low but also at moderate slope angles up to 55° (Figure 10(b)). The deposits tongue-like penetrate the horizontal plate, though their rears remain on the inclined plate.

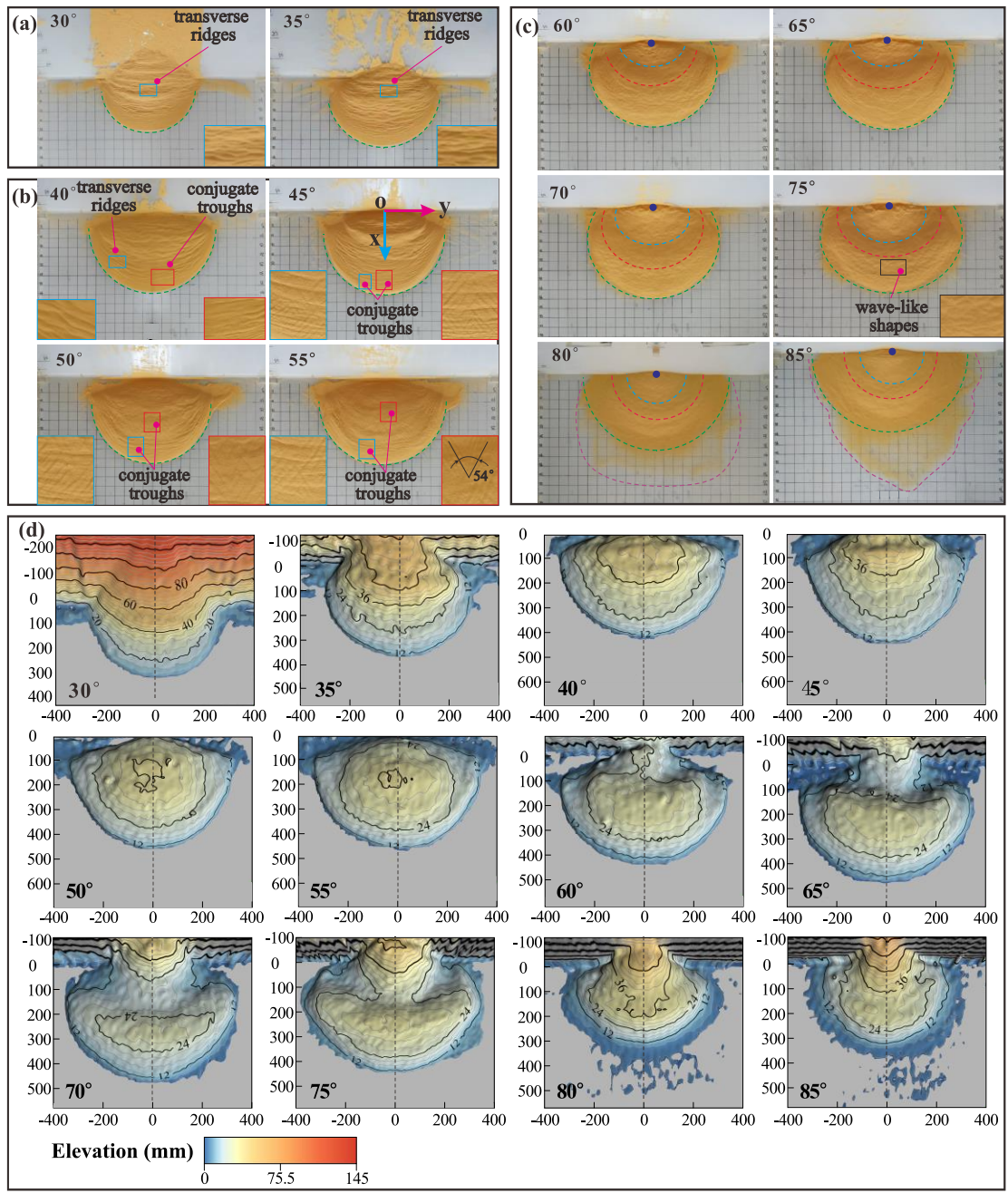

Figure 10 Surface morphologies of laboratory-landslide deposits at various slope angles: (a) low, (b) moderate, and (c) high slope angles. (d) Contour maps. The grid spacing in (a)-(c) is 5 cm. The axes in (d) indicate length in mm.

The density of transverse ridges lessens at 40°-55° (Figure 10(b)). They are now mainly observed on the deposit flanks (enlarged in Figure 11), especially at 50°-55°, and are increasingly rotated against the mean motion direction. However, washy X-shaped surface features, so-called conjugate troughs (enlarged in Figure 11), are now widely developed on the deposit surfaces, though only rudimentary at 40°. They are mainly observed on the front and centre of the deposits, but extended further and further towards the rear when the slope angle increases. The deposit boundaries are still tongue-shaped, but no longer leave remainders on the inclined plate.

At 60°-85°, neither conjugate troughs nor transverse ridges are observed on the deposit surfaces, though there now seem to be faint wave-like shapes (Figure 10(c)). Previously, these were also observed by Roche, et al. [43]. Furthermore, the deposits now exhibit two pronounced upheavals, especially at 75° (Figure 10(d)), with a vale in between. We identified a similar double-upheaval morphology in our previous study [45] and in some field data [35,46]. The front boundaries of the main deposits appear to be round. Overall, with increasing slope angle, the rear boundaries change gradually from conical to straight, while the front boundaries change gradually from tongue-like to round.

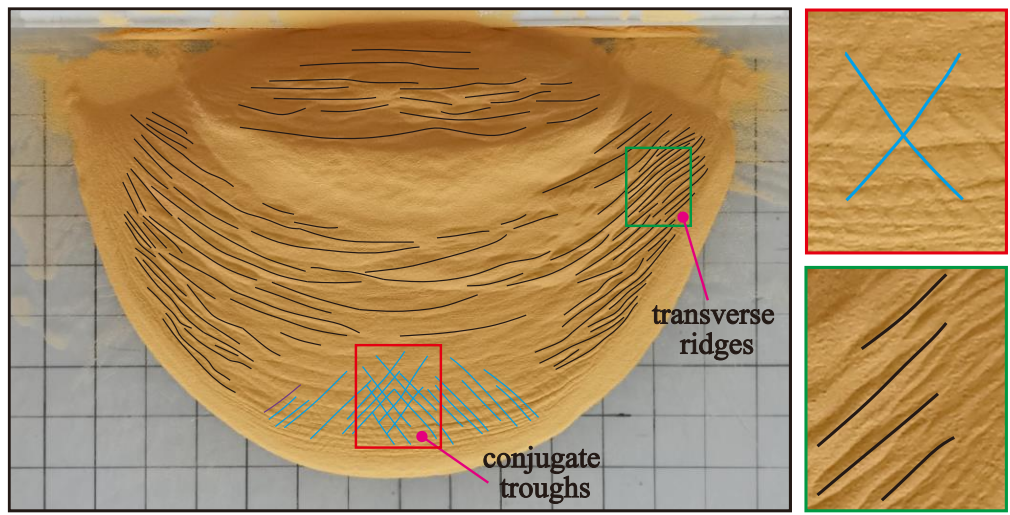

Figure 11 Transverse ridges and conjugate troughs at 45°.

### 3.2.2 Centreline morphology profiles

The centreline morphology profiles exhibit a single peak at 30°-55° (Figures 12(a) and 12(b)). The longitudinal position of the centre of gravity moves first forward and then backwards with increasing slope angle, while its vertical position decreases throughout. At 60°-85°, the profile exhibits two peaks (Figures 12(c), consistent with Figure 10(c) and 10(d), while the centre of gravity moves forward and downwards ($\theta$ =60°-70°) and then backwards and upwards ($\theta$ =70°-85°).

## 4 Discussion

### 4.1 Landslides' ranges of influence

The area occupied by a sliding mass is a measure for its range of influence[47]. Our results indicate that, without erosion of surface material along its path, the influence range of a natural landslide is strongly constrained by its initial extension, even when propagating along unrestricted terrain. In particular, its occupied length and area decrease, associated with an increase in depth and therefore basal shear stress, after reaching their maximum values near the instant of slope break.

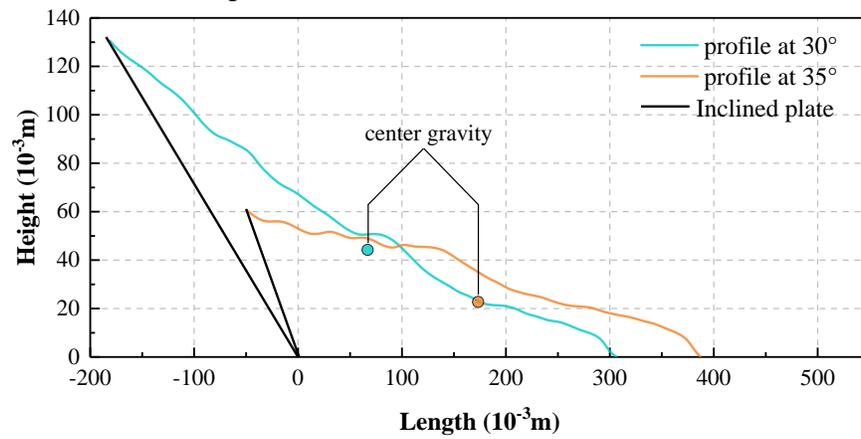

(a)

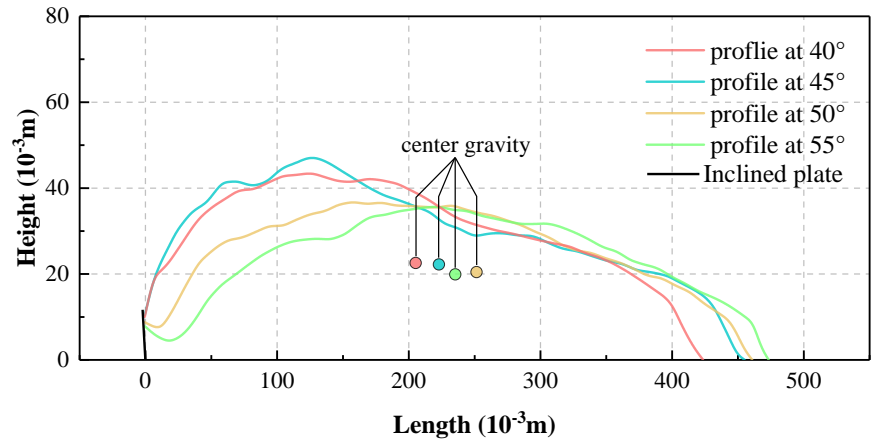

(b)

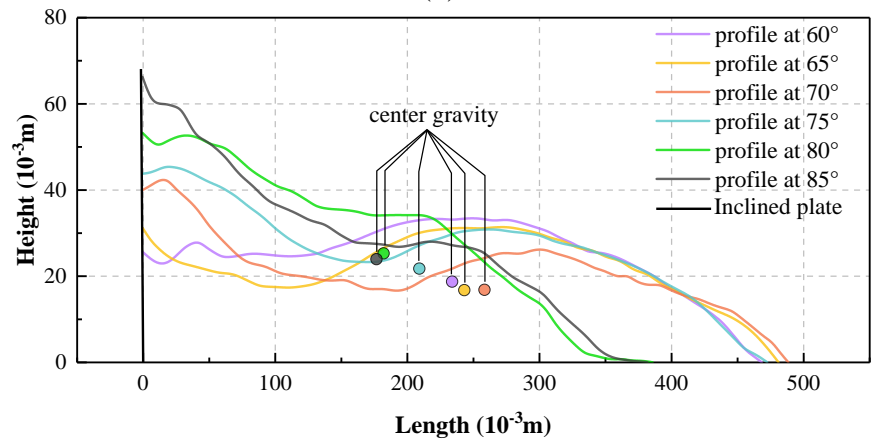

(c)

Figure 12 Centreline morphology profiles of laboratory-landslide main deposits at different slope angles (excluding the thinly spread additional deposition front at 80°-85°): (a) low, (b) moderate, and (c) high slope angles. The centres of gravity are calculated from the contour maps in Figure 10(d).

### 4.2 Physical origin of deposit morphology patterns

Transverse ridges, conjugate troughs, and double-upheavals are observed not only in our laboratory experiments but also for natural landslides [5,21,35,48]. We therefore discuss their likely physical formation origin gleaned from our experiments.

#### 4.2.1 Transverse ridges

Transverse ridges, forming in our laboratory experiments for slope angles of 30°-55° (Figure 10(a) and 10(b)), are also observed on the deposit of natural landslides, such as the Luanshibao landslide (Figure 13 (a), (b), (c), (d)) with an approximate slope angle of 45° (Tibet Plateau, Sichuan, China; Figure 13 (b) and (c)) [49]. Like in the experiments, its transverse ridges on its central portion of its deposit are almost perpendicular to its mean motion direction, but those on its flanks are rotated by an acute angle. The formation of the transverse ridges is known to be closely related to the stress state, which is similar to a thrust structure in which resistance at the front and thrust at the back act mutually on a sliding mass [18,21,25]. Their rotation and higher density on the sliding mass' flanks are due to the lower velocity magnitude and different velocity direction relative to its centre [50], causing its material on the flanks to be subject to stronger compression.

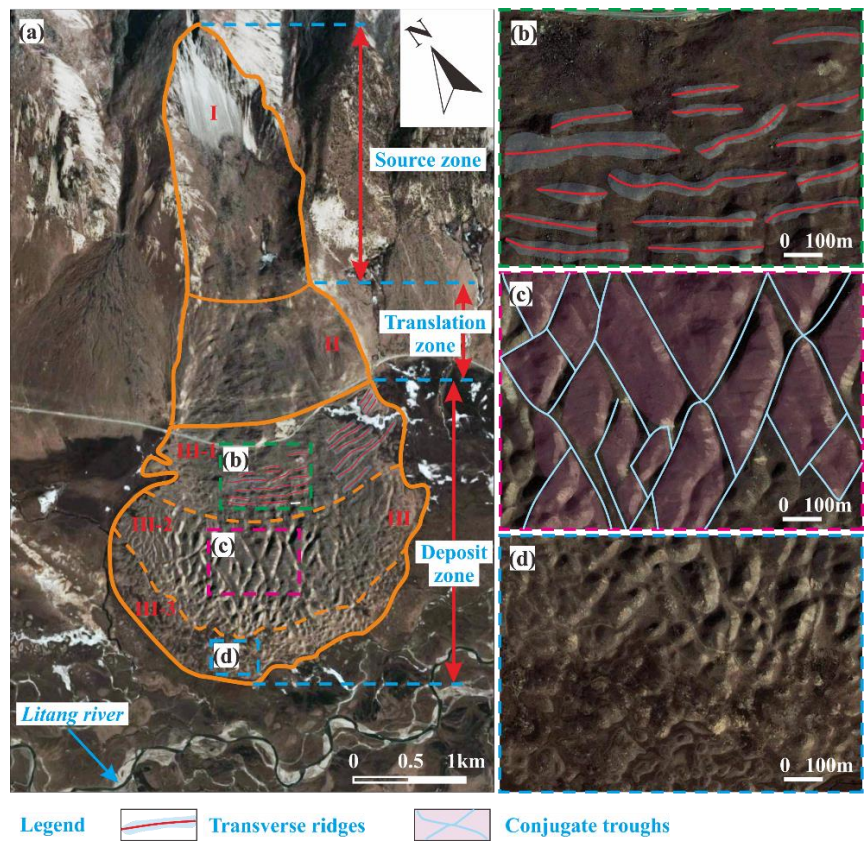

Figure 13 Morphology of the Luanshibao landslide deposit in the Tibetan Plateau, Sichuan, China. (a) Image of the Luanshibao landslide (from Google Earth); (b) transverse ridges; (c) conjugate troughs; (d) hummocks.

The physical origin of the wave-like shapes has distinct from that of the transverse ridges forming at smaller slope angles despite being morphologically similar. Generally speaking, the formation of transverse ridges is a gentle process with a comparably small landslide velocity (Figure 5(a) and Figure 14(a)) and the associated granular flow is liquid-like. In contrast, the formation of the wave-like shapes is a rapid process with a comparably large landslide velocity due to impact-induced leapfrogging of sliding mass (Figure 5(c) and Figure 14(b)) and the associated granular flow is gas-like. In addition, during the formation of transverse ridges, the Froude Number $u/\sqrt{gh}$ ($u$ is the particle velocity, and $h$ is the average height of deposit) is smaller than about 0.693, indicating that gravity plays a more important role than inertial forces. However, for the wave-like shapes, $u/\sqrt{gh}$ is larger than about 1.291, indicating that inertial forces play the dominant role.

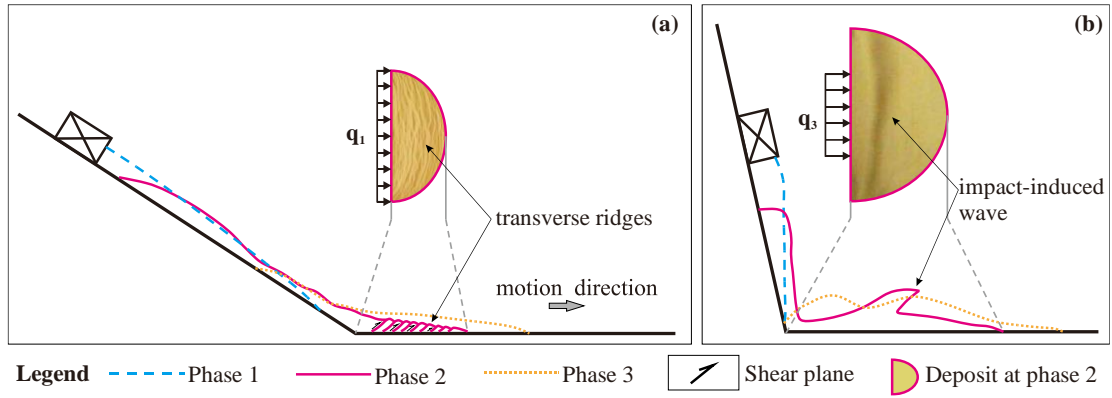

Figure 14 Schematic diagram of sliding masses at different phase for the formation of transverse ridges (a) and double upheaval (b).

### 4.2.2 Conjugate troughs

Wang, et al. [21] and Zhao, et al. [25] proposed different mechanisms for the formation of conjugate troughs following *in situ* investigations. Wang, et al. [21] suggested that conjugate troughs are formed by transport-parallel compression and radial spreading of the sliding mass. In contrast, Zhao, et al. [25] attribute their formation to a triaxial stress state of the sliding mass during motion. Based on this hypothesis, we predict the angle enclosed in the X-shaped troughs as follows: According to the Mohr-Coulomb failure criterion, granular motion occurs when a micro-unit of the deposit fails due to the triaxial shear stress $\left| \sigma_1 \text{-} \sigma_3 \right| / 2$ overcoming $\tan(\varphi) \times (\sigma_1 + \sigma_3)/2$, where $\varphi$ is the internal friction angle of the deposit material and $\sigma_1$ and $\sigma_3$ the maximum and minimum principal stresses, respectively. The direction of the shear planes just at failure is tilted by an angle of $45° - \varphi/2$ against the direction of $\sigma_1$. That is, assuming the X-shaped troughs form due to failure caused by triaxial shearing, the angle enclosed in the 'X' should be $90° - \varphi$. This prediction is consistent with the measured angles of $54°$ in our experiments ($\varphi=36°$) and $50°$ for the Luanshibao landslide [49], which consists of more resistive surface material ($\varphi=40°$, Dai, et al. [49]). Therefore, the proposition by Zhao, et al. [25] that conjugate troughs on the deposit surface form due to a triaxial shear stress is strongly supported by our experiments and the Luanshibao landslide. Note that the formation of troughs during the Luanshibao landslide is probably related to its liquified base[5,21], making it comparable to our low-frictional-base experiments. In fact, in our previous laboratory landslides along a rough base, as well as in most natural landslides, troughs did not seem to form. Furthermore, the Luanshibao landslide's curvature geometry is rather smooth and does not exhibit a sudden slope break. Such smooth geometries seem to favour trough formation at lower slope angles. The Luanshibao landslide's average slope angle was about 33° and the slope of the experiments by Shea and van Wyk de Vries [16], which also seem to exhibit troughs (their figure 8(J)), was at an average 30°.

The final deposit morphology forms due to an interplay between new emerging conjugate troughs by new stress failures and downstream-propagating older troughs (Figure 15). An available video for their formation shows in the supplementary material.

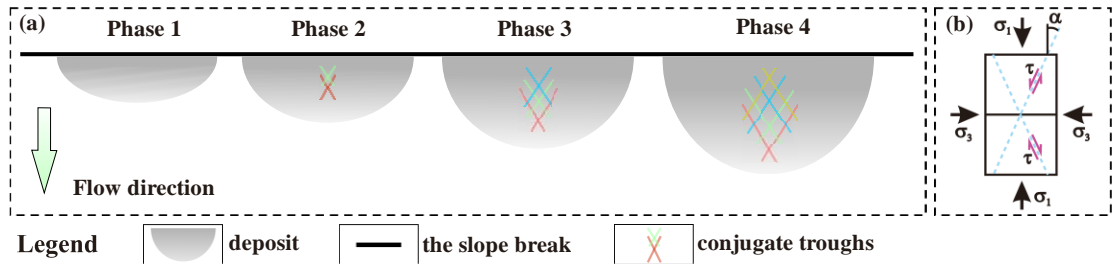

Figure 15 The formation process of conjugate troughs ($\alpha = (90° − \varphi)/2$): evolution (a) and stress condition (b) of conjugate troughs.

In addition, we hypothesize that the bumps forming on the inclined plate at moderate slope angles (Figure 5(b)), but not at low and high slope angles, and subsequently propagating onto and along the horizontal plate are the reason why conjugate troughs form. They constitute obstacles that hinder the granular flow in their wake and thereby generate sufficiently large compressive stresses in the mean flow direction for failure to occur.

### 4.2.3 Double-upheaval morphology

A multi-upheaval morphology was also observed by Roche, et al. [43] in a cylinder experiment, who rapidly released sand from a lifted cylinder that subsequently fell freely onto a flat plate. For sufficiently large fall heights, two or more circular upheavals orbiting the impact location at different radii formed. Roche, et al. [43] proposed that this was due the impact of later falling sand on the sand that had already reached the plate and formed a nearly motionless erodible surface. Once all sand was motionless, alternating crests and troughs were distributed across the surface of the deposit. We believe that our double-upheaval morphology formed due to an analogous reason, since this morphology was the more pronounced the larger the slope angle, i.e., the closer to a free fall directly onto the horizontal plate. When the rear portion of the sliding mass impacts the front portion that has already settled on the horizontal plate, the latter will surge forward and form wavy patterns (Figure 5 (c) and 14 (b)). Note that a similar wave-like surging forward of the granular material was also observed in the laboratory experiments by Mangeney, et al. [29] and Edwards and Gray [28].

## 5 Conclusions

We conducted laboratory granular flow experiments based on a physical model with an unconfined landslide geometry at a large range of slope angles. The following points are the main takeaways from these experiments:

(1) The laboratory landslides exhibit different motion characteristics at different slope angles. At low slope angles, their motion comprises three stages: uniform acceleration, constant-velocity, and deceleration. At moderate slope angles, their motion also comprises three stages: uniform acceleration, acceleration at a decreasing rate, and deceleration. At high slope angles, their motion only comprises two stages: uniform acceleration and deceleration. The runout of the landslides decreases with increasing slope angle.

(2) The length and area of the sliding masses increase first and then decrease during their whole motion. Their maximum length and area decrease with increasing slope angle. There is also a maximum landslide width. Once it is reached, the sliding masses propagate without further significant width changes.

(3) At low slope angles, transverse ridges are widely developed on the surface of the resulting deposit due to overthrusting caused by compression. At moderate slope angles, X-shaped conjugate troughs form. A Mohr-Coulomb failure model predicts the angle enclosed by the 'X' as 90°-$\varphi$, with $\varphi$ the internal friction angle, in agreement with our experiments and a natural landslide. This strongly supports the speculation by Zhao, et al. [25] that conjugate troughs form due to failure associated with a triaxial shear stress and also offers an explanation for why these patterns are observed only at moderate slope angles. At high slope angles, the deposits exhibit a double-upheaval morphology, probably because of the close similarity to free-fall regime, for which similar patterns have been observed previously [43].

**Acknowledgments**
This study would not have been possible without financial support from the Special Fund for the National Natural Science Foundation of China under Grant Nos. 42177155, 41790442, and 41702298.

**Author contributions**
Each author contributed to different parts, here listed: Conceptualisation: Yan-Bin Wu and Zhao Duan, Funding acquisition: Zhao Duan, Conducting experiments and analysis: Yan-Bin Wu, Zhao Duan, Jian-Bing Peng, and Qing Zhang; Writing: Yan-Bin Wu, Zhao Duan, and Thomas Pähtz.

**Date availability statement**
The data used to support the findings of this study are included in this paper.

**Conflicts of interest**
The authors declare that they have no known competing financial interests or personal relationships that could have appeared to influence the work reported in this paper.